\documentclass[10pt,twocolumn]{article}
\usepackage{graphicx}%
\usepackage{amsmath} 
\usepackage{amssymb}
\usepackage{bm}
\usepackage{xcolor}%
\usepackage{mystyle}
\usepackage{authblk}
\usepackage{abstract}
\newcommand{\gs}{\textsl{g}}
\newcommand{\ib}{\mathrm{i}}
\newcommand{\xb}{\mathrm{x}}
\newcommand{\retr}{\mathrm{ret}}
\newcommand{\advr}{\mathrm{adv}}
\newcommand{\Fr}{\mathrm{F}}
\newcommand{\dr}{\mathrm{dx}}
\newcommand{\dd}{\mathrm{d}}
\newcommand{\ii}{\mathrm{i}}
\newcommand{\ren}{\mathrm{ren}}

\renewcommand{\geq}{\geqslant}
\newcommand{\RC}{\mathcal{R}}
\DeclareMathOperator{\sgn}{sgn}
\DeclareMathOperator{\im}{Im}
\DeclareMathOperator{\re}{Re}
\newcommand{\bs}{\begin{subequations}}
\newcommand{\es}{\end{subequations}}

\usepackage[numbers,sort&compress]{natbib}
\definecolor{darkgreen}{rgb}{0,.5,0}
\usepackage[colorlinks,filecolor=blue,citecolor=darkgreen,unicode]{hyperref}
\title{Quest for particles production in the plane gravitational wave spacetime}
\author{Nail Khusnutdinov\thanks{email: nail.khusnutdinov@gmail.com}}
\affil{\small Centro de Matem\'atica, Computa\c{c}\~ao e Cogni\c{c}\~ao, Universidade Federal do ABC\\ \small Santo Andr\'e, 09210-170, SP, Brazil}
\date{} 

\begin{document}

\twocolumn[
\maketitle
\begin{onecolabstract}
	Massless vector and scalar Green's functions are obtained in closed form for a plane gravitational wave spacetime using three independent methods: direct solution of the Green-function equation and construction via the DeWitt and Hadamard recursive schemes. The results coincide exactly with the DeWitt--Schwinger expansion, indicating no massless particles production in this background. Contrasting interpretations, based on Bogolyubov coefficient calculations predicting massless particle creation, are also addressed.
\end{onecolabstract}
\vspace*{2em}]
\saythanks

\section{Introduction} 

The problem of particle production was first addressed in the early development of quantum electrodynamics. One of the most notable predictions in this context was made by J. Schwinger \cite{Schwinger:1951:GIVP}, who demonstrated that a strong external electric field can induce particle creation from the vacuum -- a phenomenon now known as the \textit{Schwinger effect}. In this scenario, the vacuum becomes unstable under the influence of the external field \cite{Fradkin:1991:Qeuv}, leading to the creation of particle--antiparticle pairs. Gravitational fields can likewise produce particles, a possibility first predicted in the early 1960s \cite{Parker:1968:Pceu,Zeldovich:1972:PPVPAGF} (see also the historical account by L. Parker \cite{Parker:1966:Thesis}). The most famous example of particle production by a gravitational field is Hawking radiation \cite{Hawking:1974:Bhe,Hawking:1981:Iqfbh}. Comprehensive discussions of particle creation in gravitational backgrounds can now be found in several monographs \cite{Birrell:1982:Qfcs,Fulling:1989:Aqftcst,Grib:1994:Vqesf,Parker:2009:Qftcs,Mukhanov:2007:Iqeg}.

Various methods have been proposed to compute the vacuum expectation value of the energy--momentum tensor. The most widely used is the point-splitting method \cite{Birrell:1982:Qfcs}, in which the energy--momentum tensor is expressed as the action of a two--point differential operator on the renormalized Green's function, followed by the coincidence limit. Renormalization in this approach consists of subtracting from the Green's function its DeWitt--Schwinger adiabatic expansion. Another common technique is the Pauli--Villars subtraction method \cite{Itzykson:1980:Qft}, in which each Green's function in a Feynman diagram is renormalized by subtracting analogous functions with large regulator masses, which are taken to infinity at the end of the calculation (see also the modified Pauli--Villars method in Ref.\,\cite{Vilenkin:1978:Prta}). 

In the case of flat spacetime, the vacuum state is uniquely constructed, and particle production is well-defined \cite{Fradkin:1991:Qeuv}. However, the gravitational field introduces complications to this area. The vacuum state becomes ambiguous \cite{Fulling:1973:NCFQRS}. A well-known example is the three different vacuums in Schwarzschild black hole spacetime \cite{Birrell:1982:Qfcs}, which are usually called Boulware, Hartle--Hawking, and Unruh vacuums. In each case, the annihilation and creation operators are defined with respect to a different "time", which corresponds to a different physical picture. The Boulware vacuum defines particles with respect to Schwarzschild time and exhibits no particle flux at infinity. The Hartle--Hawking vacuum is defined with respect to Kruskal coordinates and describes the thermal equilibrium of radiation with the black hole. There is particle production, but the fluxes inside and outside the black hole are equal. The Unruh vacuum describes a black hole after collapse, with a particle flux at infinity (Hawking radiation). 

The Bogolyubov method (see, e.g., \cite{Grib:1994:Vqesf,Fradkin:1991:Qeuv}) is likewise based on the concept of an unstable vacuum. The fields are assumed to be free in the infinitely distant past and future, but with different vacuum states -- commonly referred to as the \textit{in} and \textit{out} vacua. The annihilation operators that annihilate the \textit{in}  vacuum do not, in general, annihilate the \textit{out} vacuum. A linear transformation between the annihilation and creation operators of the \textit{in} and \textit{out} vacua involves two parameters, known as the Bogolyubov coefficients. One of these coefficients characterizes scattering, while the other determines the density of created particles. In principle, the Bogolyubov method and the vacuum expectation value approach should yield identical results; however, this equivalence does not hold in the case of a plane gravitational wave background.

The interaction between gravitational waves (GWs) and electromagnetic waves (EMWs) has a long history. Within classical physics, the conversion of weak GWs to EMWs -- and vice versa -- in the presence of external electric or magnetic fields was first considered by Gertsenshtein in 1962 \cite{Gertsenshtein:1962:WRLGW} and later in Refs.\,\cite{Boccaletti:1970:Cpgvvsef,Zeldovich:1974:Egwsmf,Dolgov:2023:GPCCSEMF}. In the framework of quantum gravity in the weak--field approximation \cite{Bronstein:2012:RQtwgf,Gupta:1952:QEGFGT}, the cross sections of such processes have been calculated (see, e.g., \cite{Mitskevich:1969:PFGRTR}); the effect is proportional to the square of the external electric or magnetic field strength. The conversion of weak GWs to EMWs in the absence of external electromagnetic fields has been investigated in Refs.\,\cite{Tourrenc:1970:Epdg,Skobelev:1975:GPI}, with the corresponding cross section scaling as $(G/c^3)^2$, where $G$ is the gravitational constant and $c$ is the speed of light.

The Green function for a scalar field in a plane gravitational-wave spacetime was first derived by Gibbons \cite{Gibbons:1975:Qfpps}, who demonstrated that it coincides exactly with its DeWitt--Schwinger expansion. This result implies that plane gravitational waves do not induce scalar particle production. In contrast, Refs.\,\cite{Jones:2017:Ppgwb,Jones:2018:Sfvevigwb,Jones:2019:Igrer} present a different scenario for massless scalar fields, suggesting that particle production may occur when the particle momenta are exactly aligned with the propagation direction of the gravitational wave. This claim is based on the analysis of the Bogolyubov coefficient $\beta$, which quantifies the particle number density. The Bogolyubov approach, originally developed by Garriga and Verdaguer \cite{Garriga:1991:Sqpgpwa}, yields $\beta = 0$ for massive particles, confirming the results of Ref.\,\cite{Gibbons:1975:Qfpps}. However, Refs.\,\cite{Jones:2017:Ppgwb,Jones:2018:Sfvevigwb,Jones:2019:Igrer} argue that in the massless case, $\beta$ can be nonzero when the gravitational-wave and particle momenta are precisely collinear. Specifically, $\beta$ is proportional to $\delta(k_v + l_v)$, where $k_v = \omega - k_x$ and $l_v = \omega' - l_x$ are the $v$-components of the wave vectors of the \textit{in} and \textit{out} scalar modes\footnote{$\omega$ and $\omega'$ are frequencies of \textit{in} and \textit{out} scalar waves, correspondingly, and the direction of propagation of the gravitational wave is the axis $x$.}. For massive particles, $k_v + l_v > 0$, ensuring $\beta = 0$; for massless particles, the sum can vanish, allowing $\beta \neq 0$. This leads to an apparent tension between the DeWitt--Schwinger and Bogolyubov approaches in the context of plane gravitational wave backgrounds (for a comparative discussion, see Refs.\,\cite{Birrell:1982:Qfcs,Fulling:1989:Aqftcst}).

In this work, we compute the Green function for the vector (Maxwell) field using three complementary methods: direct integration of the Maxwell equations, the DeWitt--Schwinger expansion, and the Hadamard elementary solution. The direct method is based on the definition of the Green's function as a specific solution to the corresponding operator with a delta-like current and boundary conditions. The well-known method uses the full, orthogonal set of eigenfunctions of the operator. We define the boundary conditions with respect to the retarded time, $u$. The DeWitt method \cite{DeWitt:1965:Dtgf} represents the Green function as an asymptotic series over the variable $s \ll \sigma(x,x')$, where $\sigma (x,x')$ is the Synge's "world function"\/ \cite{Synge:1976:Rgt}, which is half the square distance between two events. This series is valid only locally, for small distances between events. In general, the exact Green's function does not coincide with the DeWitt expansion, though their ultraviolet  singularities do. The main application of this series is the DeWitt-Schwinger renormalization in a curved background \cite{Birrell:1982:Qfcs}. The coincidence limit of the difference between the exact Green function and its DeWitt expansion is understood as a quantum contribution and the manifestation of the quantum origin of fields. The Hadamard approach is based on the general structure of Green function singularities in different dimensions. In even dimensions, there are pole and logarithmic singularities, and in odd dimensions, there are only poles. Local series expansions over $\sigma (x,x')$ exist for the factors of poles and logarithms, similar to the DeWitt series. The application of these series is similar to the DeWitt series  \cite{Birrell:1982:Qfcs}. General relations between different Green functions can be found in DeWitt's book \cite{DeWitt:1965:Dtgf}.

We show that the retarded Green function for the vector potential has support not only on the past light cone but also within its interior. By contrast, the Green function of the physical electromagnetic field (the field-strength tensor) is supported strictly on the past light cone. This property, in the context of the Huygens principle, was previously noted in Ref.\,\cite{Kunzle:1968:Mfshp} for fields of (0,1,2)-form type, and for the Maxwell field in Ref.\,\cite{Khusnutdinov:2020:SHp} in the context of self-force calculations.

The paper is organized as follows. Sec. \ref{sec:geometry} reviews the general properties of plane gravitational waves and establishes a convenient computational frame. Sec. \ref{sec:Greens} presents the derivation of the scalar and massless vector Green functions. Sec. \ref{sec:conclusion} discusses the results and their physical implications.

\section{The geometry, biscalar, bivector and frame} \label{sec:geometry}

Let us first consider the general form of a plane gravitational wave propagating in the $x$-direction. The corresponding metric\footnote{Hereafter, $i,j=2,3=y,z$,  $u,v = (t \mp x)/\sqrt{2}$, $\mu,\nu = u,v,y,z$,  and $\sqrt{-\gs} =  L^2$. The overdot is derivative with respect to $u$. $(a),(b) =0,1,2,3$ are frame indexes, $\xb = (u,v,y,z)$ -- event in spacetime and $\dr = \sqrt{-\gs(\xb)} \dd^4\xb$ is the invariant measure. The units with $c=G=\hbar =1$ are used.} can be written as (see, e.g., \cite{Stephani:2006:Esefe}) 
\begin{equation}\label{eq:metric}
	\dd s^2 = - 2\dd u \dd v + \gamma_{ij}(u) \dd x^i \dd x^j . 
\end{equation}
The coordinates in which the metric takes the form \eqref{eq:metric}  are known as Baldwin--Jeffery--Rosen (BJR) coordinates, named after the works \cite{Baldwin:1926:RTPW,Rosen:1937:Ppwgtr} in which this representation first appeared. It is well established that the spacetime geometry of a plane gravitational wave can also be expressed in a single chart using Brinkmann coordinates \cite{Brinkmann:1923:RSCES}, where the metric takes the form 
\begin{equation}\label{eq:metric2}
	\dd s^2 = -2 \dd U \dd V + \delta_{ij} \dd X^i \dd X^j + K_{ij}(U) X^i X^j \dd U^2.
\end{equation} 
The transformation from the Brinkmann form \eqref{eq:metric2} to the BJR \eqref{eq:metric}  is given in Refs.\,\cite{Gibbons:1975:Qfpps,Duval:2017:csopgw} by 
\begin{equation}\label{eq:transform}
	U=u, X^i = P^i_k x^k, V = v + \frac{1}{4} \dot \gamma_{ik} x^i x^k, 
\end{equation}
where the matrix $P^i_k$ satisfies the equations 
\begin{equation}
	\gamma_{ik} = P^n_i P^n_k,\ \ddot{P}^i_k = K_{il} P^l_k,\ \dot{P}^n_i P^n_k = \dot{P}^n_k P^n_i.  
\end{equation}

A gravitational wave sandwich spacetime can be covered by a single coordinate chart in the Brinkmann representation, but requires two charts in the BJR representation. In BJR coordinates, a coordinate singularity appears after the wave exits the sandwich region. In Brinkmann coordinates, the relation  $\RC^i_{\cdot\ uju} = K_{ij}$ holds, and a sandwich of thickness $T = u_2 - u_1 > 0$ is characterized by $K_{ij} = 0$ for $u < u_1$ and $u > u_2$. Outside the sandwich, the spacetime is Minkowskian. This is not the case in the BJR frame: after the wave passes, one has $\ddot{\mathbf{P}} = 0$ which implies $\mathbf{P} = \mathbf{A} + u \mathbf{B}$, where $\mathbf{A}$ and $\mathbf{B}$ are constant matrices satisfying $\mathbf{B}^T \mathbf{A} = \mathbf{A}^T\mathbf{B}$  with $\mathbf{B}\not =0$ \cite{Garriga:1991:Sqpgpwa}. 

The metric \eqref{eq:metric} is invariant under a 5-parameter symmetry group $G_5$ \cite{Bondi:1959:GwgrIEpw}, generated by the Killing vectors $\bm{\xi}_1 = \partial_v, \bm{\xi}_i = \partial_i$   and $\bm{\xi}_4 = y \bm{\xi}_1 + p^{2i}(u) \bm{\xi}_i, \bm{\xi}_5 = z \bm{\xi}_1 + p^{3i}(u) \bm{\xi}_i$ where the matrix $p_{ij}$ has the form \begin{equation}\label{eq:p}
	p^{ij} = \int_{u'}^{u} \gamma^{ij}\dd u,\ p_{ij}p^{jk} = \delta^k_i. 
\end{equation} 
The vector $\bm{\xi}_1$ is a covariantly constant null vector, satisfying $\nabla_\alpha\bm{\xi}_1=0$. The first three Killing vectors generate an Abelian subgroup $G_3$, which acts transitively on the null hypersurfaces $u=const$. The remaining two vectors may be interpreted as Carroll boosts without rotation \cite{Duval:2017:csopgw}  (see also \cite{Bondi:1959:GwgrIEpw}). For the metric \eqref{eq:metric}, all polynomial curvature invariants vanish, $\RC = \RC_{\mu\tau} \RC^{\mu\tau} = \RC_{\alpha \beta \mu\tau} \RC^{\alpha \beta \mu\tau} = 0$.

Taking into account the three Killing vectors of the subgroup $G_3$, we obtain the first integrals of the geodesic equations: $u_i = u_i^0, u_v = u_v^0$ and, from the normalization condition, the final integral $u_u = ( 1 + \gamma^{ij}u_i^0 u_j^0 )/2 u_v^0$. Integration of these relations yields the world function $\sigma$ (i.e., half the squared geodesic interval between the points), 
 \begin{equation}
	\sigma (\xb,\xb') = \delta u \left( -\delta v + \frac{1}{2}\delta x^i \delta x^j p_{ij} \right),  
\end{equation} 
where $\delta x^\mu = x^\mu - x'^\mu$. The van Vleck--Morette determinant is given by $\Delta =\frac{ \delta u^2 \det p_{ij}}{L^2 L(u')^2}$.

To determine the bivector $\gs_{\mu\nu'}$ of parallel transport along the geodesic, we solve the parallel transport equations
$u^\nu \omega_{\mu; \nu}=0$  for a vector $\omega_\mu$, and express the solution as $\omega_\mu (u) = \gs_{\mu\nu'} \omega^{\nu'}(u')$. For a convenient representation of the bivector, we employ the frame (row index $a$ denoting the vector number) 
\begin{equation}
	\lambda^\mu_{(a)}  = 
	\begin{pmatrix}
		1&0&0&0\\
		0&1&0&0\\
		0&0&\frac{L^2 \cos \psi  - \gamma_{yz} \sin\psi}{\sqrt{\gamma_{yy}}L^2}& \frac{\sqrt{\gamma_{yy}} \sin\psi}{L^2}\\
		0&0&-\frac{\gamma_{yz}\cos\psi + L^2\sin \psi }{\sqrt{\gamma_{yy}}L^2}& \frac{\sqrt{\gamma_{yy}} \cos\psi}{L^2}
	\end{pmatrix},
\end{equation}
where $\psi$ is an arbitrary function of $u$ and $u'$. The standard orthonormality conditions are satisfied, 
\begin{equation}
	\lambda^\mu_{(a)} \lambda^\nu_{(b)} \eta^{ab} = \gs^{\mu\nu},\ \lambda^\mu_{(a)} \lambda^\nu_{(b)} \gs_{\mu\nu} = \eta_{ab},
\end{equation} 
with 
\begin{equation}
	\eta_{ab} =
	\begin{pmatrix}
		0&-1&0&0\\
		-1&0&0&0\\
		0&0&1&0\\
		0&0&0&1
	\end{pmatrix}.
\end{equation} 

By setting 
\begin{equation}
	\psi = \int^u_{u'} \frac{\gamma_{yz} \dot{\gamma}_{yy} - \dot{\gamma}_{yz} \gamma_{yy}}{2\gamma_{yy} L^2}\dd u,
\end{equation} 
we obtain the frame components of the bivector of parallel transport (see Appendix \ref{sec:appA}) 
\begin{equation}\label{eq:gab}
	\gs_{(a)(b')} = 
	\begin{pmatrix}
		-\frac{1}{2} \delta \upsilon_{(a)} \delta \upsilon^{(a)} & -1 & \delta \upsilon_{(2)} & \delta \upsilon_{(3)}\\
		-1& 0 & 0 & 0 \\
		-\delta \upsilon_{(2)} & 0 & 1 & 0 \\
		-\delta \upsilon_{(3)} & 0 & 0 & 1
	\end{pmatrix},
\end{equation}
where the index $a$ labels the row and $b$ the column. We introduce the vector $\upsilon_\mu = \left(0,0,u_y^0/u_v^0,u_z^0/u_v^0\right)$, where $u_\mu$ is the tangent vector to the geodesic, with frame components  $\upsilon_{(a)} = \upsilon_\mu \lambda^\mu_{(a)}$,  $\upsilon_{(b')} = \upsilon_{\mu'} \lambda^{\mu'}_{(b')}$. The difference $\delta \upsilon_{(a)} = \upsilon_{\mu} \delta \lambda^\mu_{(a)}$, where $\delta \lambda^\mu_{(a)} = \lambda^\mu_{(a)}(u)  - \lambda^{\mu}_{(a)}(u')$. For the geodesic we have $u_i^0/u_v^0 = - p_{ij} \delta x^j$. The symmetry condition $\gs_{(a)(b')}(\xb,\xb') = \gs_{(b')(a)} (\xb',\xb)$ is satisfied. 

With this choice of $\psi$, the Ricci tensor can be expressed in the form 
\begin{equation}\label{eq:Rmn}
\RC_{\mu\tau} = - \delta_\mu^u \delta_\tau^u \left((\ln L^2)\, \ddot{} + \gamma_{ij} \dot{\lambda}^i_{(a)} \dot{\lambda}^{j(a)}\right).
\end{equation}  

\section{The Green functions} \label{sec:Greens}

\subsection{The scalar Green functions}

In this section, we present a slightly more detailed derivation of the scalar Green's function, first performed by Gibbons \cite{Gibbons:1975:Qfpps}. We calculate Green functions using the eigenvalues of the corresponding operator. The same result is demonstrated in the framework of the DeWitt series expansion and the Hadamard elementary solution. 

We first consider the scalar Green function satisfying the equation 
\begin{equation}\label{eq:scalarE}
	(\Box  - m^2) D = - \frac{\delta^{(4)}(\xb - \xb')}{\sqrt{-\gs}}, 
\end{equation}
where $\Box = L^{-2} \partial_\mu \left(L^2 \gs^{\mu\tau} \partial_\tau\right)$ is the Laplace--Beltrami operator. Using the set of functions \cite{Gibbons:1975:Qfpps}
\begin{equation}\label{eq:fullset}
	\phi_k = \frac{1}{4\pi^2 L} e^{\ib k_\mu \xb^\mu + \frac{\ib }{2k_v}  k_i k_j \int^u \gamma^{ij}\dd u},
\end{equation}
which solve the eigenvalue problem 
\begin{equation}
	(\Box  - m^2)\phi_k = (2k_v k_u - m^2)\phi_k,
\end{equation} 
we obtain the scalar Green function 
\begin{equation}\label{eq:D1}
	D =  -\int \frac{\phi_k (\xb) \phi^*_k (\xb')}{2k_v k_u  - m^2}  \dd^4k.
\end{equation} 
The set of functions \eqref{eq:fullset} forms a complete and orthogonal basis: 
\begin{align*}
	\int L^2\dd^4\xb \phi_k(\xb) \phi_{k'}(\xb) &= \delta^{(4)}(k-k'),\\
	\int \dd^4 k \phi_k(\xb) \phi_k(\xb') &= \frac{\delta^{(4)}(\xb - \xb')}{L^2}.
\end{align*}

Expression\eqref{eq:D1} is ill-defined as it stands; an infinitesimal imaginary term must be added to the denominator, thereby specifying the integration contour. In other words, one must first select the variable with respect to which the causal structure -- future or past -- is defined. The natural choice is the coordinate $u$, associated with the covariantly constant Killing vector $\bm{\xi}_1$.

With this choice, the retarded and advanced Green functions are given by 
\begin{equation}\label{eq:scalarD}
	D_{\retr/\advr} =  -\int \frac{ \phi_k (\xb) \phi^*_k (\xb')\dd^4k}{2k_v (k_u \mp \ib 0)  - m^2}.
\end{equation} 
The Feynman Green function is obtained in the usual way by adding a negative imaginary part to the mass term \begin{equation}
	D_\Fr = -\int \frac{\phi_k (\xb) \phi^*_k (\xb')\dd^4k}{2k_v k_u - m^2 + \ib 0}.  
\end{equation}

Employing the prescription $k_v \to k_v - \ib 0 \sgn (\delta u)$ for convergence, 
\begin{equation}\label{eq:Gauss}
	\int \dd^2 k e^{\ib k_i \delta \xb^i + \frac{\ib k_i k_l p^{il}}{2k_v}} = \frac{2 \pi \ib k_v \sgn (\delta u)}{\sqrt{\det p^{il}}}e^{- \frac{\ib k_v}{2} p_{il} \delta \xb^i \delta \xb^l},
\end{equation} 
we recover in the massless case the expression 
\begin{equation}\label{eq:dret}
	D_{\retr/ \advr} = \frac{\Delta^{\frac 12}}{4\pi} \theta (\pm \delta u) \delta (\sigma), D_\Fr = \frac{\ib}{8\pi^2} \frac{\Delta^{\frac{1}{2}}}{\sigma + \ib 0}.
\end{equation}
This result was first obtained by Gibbons in Ref.\,\cite{Gibbons:1975:Qfpps}. As expected, $\re D_\Fr = \overline{D} = (D_\retr + D_\advr)/2$. The Hadamard elementary function $D^{(1)}$ is given by 
\begin{equation}
	D^{(1)} = 2 \im D_\Fr = \frac{\Delta^{\frac{1}{2}}}{4\pi^2} \frac{\mathcal{P}}{\sigma}.
\end{equation}

The same expressions can be obtained via the series expansion method of DeWitt \cite{DeWitt:1965:Dtgf} 
\begin{displaymath}
	\overline{D} = \frac{\Delta^{\frac 12}}{8\pi} \delta (\sigma) +  \frac{\Delta^{\frac 12}}{8\pi} \theta (-\sigma) \left( a^{(1)} - \frac{\sigma}{2} a^{(2)} + \ldots\right),
\end{displaymath}
where the biscalars (heat kernel coefficients) $a^{(n)}$ are solutions of the chain 
\begin{displaymath}
	n a^{(n)} + \sigma^{;\alpha} a^{(n)} = \Delta^{-\frac 12} (\Delta^{\frac 12} a^{(n-1)})_{;\alpha}^{\ \ ;\alpha},
\end{displaymath}
with $a^{(0)} = 1$. Since  $(\Delta^{\frac 12})_{;\alpha}^{\ \ ;\alpha} = 0$ , it follows that all heat kernel coefficients $a^{(n)}(x, x') = 0$ for $n \geq 1$, and consequently, the Green functions take the form given in Eq. \eqref{eq:dret}. 

An identical result is obtained by analytically continuing the Hadamard elementary solution  \cite{Adler:1977:Rstvsppgbm}
\begin{equation}
	D_\Fr = \frac{\ib \Delta^{\frac{1}{2}}}{(4\pi)^2} \left(\frac{2}{\sigma + \ib 0} + v \ln (\sigma + \ib 0) + w\right),
\end{equation}
where the biscalars $v$ and $w$ are expressed as series expansions 
\begin{equation}
	v = \sum_{n=0}^\infty v^{(n)} \sigma^n,\ w = \sum_{n=0}^\infty w^{(n)} \sigma^n.
\end{equation}
Owing to $\Delta_{;\alpha}^{\ ;\alpha} =0$, the first equation of the chain 
\begin{equation}
	v^{(0)} + \sigma^{;\mu} v^{(0)}_{;\mu} = - \Delta^{-\frac{1}{2}} (\Delta^{\frac{1}{2}})_{;\alpha}^{\ ;\alpha} = 0,
\end{equation}
has the solution $v^{(0)} = 0$, while the remaining equations ($n\geq 1$) 
\begin{equation}
	v^{(n)} + \frac{1}{n+1}\sigma^{;\mu} v^{(n)}_{;\mu} = - \frac{1}{2n (n+1)}\Delta^{-\frac{1}{2}} (\Delta^{\frac{1}{2}} v^{(n-1)} )_{;\alpha}^{\ ;\alpha},
\end{equation}
yield $v^{(n)} =0$. Hence, $v = 0$ throughout. The chain for $w^{(n)}$  has the same structure, but, as is well-known, $w^{(0)}$ remains arbitrary. Its value is determined by the boundary conditions \cite{Adler:1977:Rstvsppgbm}. Using the mode expansion, one finds $w^{(0)} = w = 0$ (see Eq.  \eqref{eq:dret}).

\subsection{The vector Green functions}

We begin by calculating the retarded and advanced Green functions through a direct solution of the Maxwell equations using the eigenfunctions of the scalar operator. Next, we derive the same expressions using the DeWitt series expansion and the Hadamard elementary solution. 

In the covariant Lorentz gauge, $A^\mu_{;\mu} = 0$, the Maxwell equations take the form 
\begin{equation}
	A_{\mu\ \ ;\nu}^{\ \ ;\nu} - \RC_\mu^{\ \ \nu} A_\nu = -4\pi J_\mu. 
\end{equation}
From these equations, we obtain the relations for the frame components 
\begin{equation}
	\Box A_{(a)} = A_{(b)} \left(\RC_{(a)}^{\ \ (b)} - \lambda^\mu_{(a)} \lambda^{(b);\nu}_{\mu\ \ \ \ ;\nu} \right)- 2 A_{(b),\nu} \lambda^{(b);\nu}_\mu \lambda^\mu_{(a)} - 4\pi J_{(a)},
\end{equation}
or, equivalently, in manifest form 
\bs
\begin{align}
	\Box A_{(0)}&= A_{(1)}  (\ln L^2)\, \ddot{} - 2\dot{\lambda}^{(k)i} \partial_i A_{(k)}  - 4\pi J_{(0)},\label{eq:M1} \\
	\Box A_{(1)}&= - 4\pi J_{(1)},\label{eq:M2}\\
	\Box A_{(l)}&= - 2\dot{\lambda}_{(l)}^i \partial_i A_{(1)}  - 4\pi J_{(l)}, \label{eq:M3}
\end{align}
\es
where $\Box = L^{-2} \partial_\mu \left(L^2 \gs^{\mu\tau} \partial_\tau\right)$ is the Laplace--Beltrami operator. Solving these equations, we express the solution as 
\begin{equation}
	A_{(a)}(\xb) =  4\pi \int D_{(a)(b)} (\xb, \xb') J^{(b)}(\xb') \dr',
\end{equation}
where $D_{(a)(b)} = D_{\mu\nu'}\lambda^\mu_{(a)}(\xb) \lambda^{\nu'}_{(b)} (\xb')$ are the frame components of the vector Green function.

Consider first Eq. \eqref{eq:M2}. Its particularly simple form follows from the existence of a covariantly constant Killing vector $\bm{\xi}_1$. Using the retarded scalar Green function \eqref{eq:scalarD}, corresponding to Eq. \eqref{eq:scalarE}, we obtain 
\begin{equation}\label{eq:A1}
	A_{(1)} =  4\pi \int D_\retr (\xb, \xb') J_{(1)}(\xb') \dr'.
\end{equation}
and therefore$D_{(1)(0)}^\retr = -D_{\retr} = \gs_{(1)(0)} D_{\retr}$.

From Eq. \eqref{eq:A1}, the retarded solutions of Eqs. \eqref{eq:M3} take the form 
\begin{equation*}
	A_{(i)} = 4\pi \int \left(D_\retr (\xb; \xb') J_{(i)}(\xb')  + G_{(i)}(\xb;\xb') J_{(1)}(\xb')\right)\dr',
\end{equation*}
where
\begin{equation}
	G_{(i)}(\xb;\xb') = 2\int D_\retr (\xb;\xb'') \dot{\lambda}_{(i)}^{j''} \partial_{j''} D_\retr (\xb'';\xb') \dr''.
\end{equation}
		
Employing the mode representation \eqref{eq:scalarD}  for $ D_\retr$ and integrating over $v'', y'', z''$, we arrive at 
\begin{align}
	G_{(i)} &= \frac{1}{(2\pi)^5 L(u) L(u')} \int \dd u'' \int \frac{\dd^4k}{2k_v (k_u - \ib 0)} \frac{\dd k_u' }{k_v (k_u' - \ib 0)} \nonumber \\
	&\times e^{\ib k_u (u-u'') + \ib k_u' (u'' - u) + \ib k_v \delta v + \ib k_l \delta \xb^l + \frac{\ib k_i k_l p^{il}}{2k_v} }  \ib k_{j''}  \dot{\lambda}_{(i)}^{j''} . 
\end{align}
By invoking the standard integral representation of the step function $\theta(t)$, it follows immediately that 
\begin{align}
	&\int \dd u''  \int \frac{\dd k_u}{k_u - \ib 0} \frac{\dd k_u' }{k_u' - \ib 0} e^{\ib k_u (u-u'') + \ib k_u' (u'' - u)}\nonumber \\ 
	&= 2\pi \ib \int_{u'}^u \dd u''  \int \frac{\dd k_u}{k_u - \ib 0}e^{\ib k_u \delta u},
\end{align}
and hence 
\begin{equation}
	G_{(i)} = - \int \frac{\dd^4k\ \phi_k (\xb) \phi^*_k (\xb')}{2k_v^2 (k_u - \ib 0)}  k_{j''} \int_{u'}^u \dd u''  \dot{\lambda}_{(i)}^{j''}. 
\end{equation} 

Using integral \eqref{eq:Gauss}, we find  $G_{(i)} = -D_{\retr} \delta \upsilon_{(i)}$, which implies 
\begin{equation}
	D_{(i)(i)}^\retr = D_{\retr} = \gs_{(i)(i)} D_{\retr}, \ D_{(i)(0)}^\retr =  \gs_{(i)(0)}D_{\retr}. 
\end{equation}

With the solutions for $A_{(1)}$ and $A_{(i)}$ in hand, we can solve the first equation \eqref{eq:M1} in the same manner. To demonstrate the emergence of the tail term, we focus on the first term on the right-hand side of Eq. \eqref{eq:M1}. Its contribution to the Green function is given by \begin{equation}
	4\pi \int Q_{(1)} (\xb; \xb') J_{(1)}(\xb') \dr',
\end{equation}
where 
\begin{equation}
	Q_{(1)} = -\int D_\retr (\xb;\xb'')  D_\retr (\xb'';\xb') \partial^2_{u'' u''}(\ln L^2) \dr ''.
\end{equation}

Integrating over $v'', y'', z''$, and subsequently over $k_u, k_u', k_y, k_z$, we obtain 
\begin{equation}
	Q_{(1)} = \frac{\ib \theta (\delta u) \Delta^{\frac{1}{2}}}{(2\pi)^2 \delta u} \delta( \ln L^2)\dot{} \int \frac{\dd k_v}{4k_v} e^{\ib k_v (\delta v - \frac{1 }{2} p_{il} \delta \xb^i \delta \xb^l)}. 
\end{equation} 
As noted earlier, the integral \eqref{eq:Gauss} is convergent provided $\im k_v <0$, i.e., $k_v \to k_v - \ib 0$. With this prescription, we obtain \begin{equation}
	Q_{(1)} = -\frac{\Delta^{\frac{1}{2}}}{8\pi \delta u} \delta( \ln L^2)\dot{} \theta (\delta u)  \theta(-\sigma).
\end{equation}
and finally arrive at the expression 
\begin{equation}\label{eq:Green}
	D_{\mu\tau'}^\retr = \gs_{\mu\tau'}D_\retr + \theta (\delta u)\theta (-\sigma) h_{\mu\tau'},  
\end{equation}
where 
\begin{equation}\label{eq:tailw}
	h_{\mu\tau'} = \delta_\mu^u \delta_{\tau'}^{u'} \frac{\Delta^{\frac 12}}{8\pi \delta u}  \left(\delta( \ln L^2)\dot{} + p_{ij} \delta \lambda^i_{(a)} \delta \lambda^{j(a)} \right).
\end{equation}

We now employ the series expansion method of DeWitt  \cite{DeWitt:1965:Dtgf}
\begin{displaymath}
	\overline{D}_{\mu\tau'} = \frac{\Delta^{\frac 12}}{8\pi} \gs_{\mu\tau'} \delta (\sigma) +  \frac{\Delta^{\frac 12}}{16\pi} \theta (-\sigma) \left( a^{(1)}_{\mu\tau'} - \frac{\sigma}{2} a^{(2)}_{\mu\tau'} + \ldots\right),
\end{displaymath}
to obtain the real part of the Feynman Green function,  $\re D^\Fr_{\mu\tau'} = \overline{D}_{\mu\tau'} = (D_{\mu\tau'}^\retr + D_{\mu\tau'}^\advr)/2$. In this approach, the bivectors $a^{(n)}_{\mu\tau'}$ satisfy the chain of equations (for $n \geq 1$) 
\begin{equation}\label{eq:an}
	n a^{(n)}_{\mu\tau'} + \sigma^{;\alpha} a^{(n)}_{\mu\tau';\alpha} = \Delta^{-\frac 12} (\Delta^{\frac 12} a^{(n-1)}_{\mu\tau'})_{;\alpha}^{\ \ ;\alpha} - \RC_\mu^{\ \ \alpha} a^{(n-1)}_{\alpha\tau'},
\end{equation}
with $a^{(0)}_{\mu\tau'} = \gs_{\mu\tau'}$. 

For the frame components, we obtain the relations 
\begin{align}
	&(n a^{(n)}_{\mu\tau'} + \sigma^{;\alpha} a^{(n)}_{\mu\tau';\alpha}) \lambda^\mu_{(a)} \lambda^{\tau'}_{(b')} = \Delta^{-\frac{1}{2}}\left(\Delta^{\frac{1}{2}} a^{(n-1)}_{(ab')}\right)_{;\alpha}^{\ ;\alpha}\nonumber\\
	&+ 2  \Delta^{-\frac{1}{2}} \left(\Delta^{\frac{1}{2}} a^{(n-1)}_{(cb')}\right)_{;\alpha} \lambda^{(c);\alpha}_\mu \lambda^\mu_{(a)} + a^{(n-1)}_{(cb')} \left(\lambda^{(c);\alpha}_{\mu;\alpha} \lambda^\mu_{(a)} - \RC_{(a)}^{\ (c)} \right).
\end{align}

Let us first consider the case $n = 1$, with $a^{(0)}_{(ab')} = \gs_{(ab')}$. In what follows, it is important to note that  $\gs_{(ab')}$does not depend on the coordinate $v$, $\gs_{(00')}$ is quadratic in the coordinate differences $\delta x^i$, and $\gs_{(0i')}, \gs_{(i0')}$ are linear in $\delta \xb^i$. All other components are constants (see Eq. \eqref{eq:gab}). Taking this into account, and using definition  \eqref{eq:p}, we obtain 
\begin{align}
	\Delta^{-\frac{1}{2}}\left(\Delta^{\frac{1}{2}} \gs_{(ab')}\right)_{;\alpha}^{\ ;\alpha}  &= \ \dot{p}_{ij} \delta \lambda_{(c)}^i \delta \lambda^{j(c)} \delta_a^0 \delta_{b'}^{0'}, \nonumber \\
	2  \Delta^{-\frac{1}{2}} \left(\Delta^{\frac{1}{2}} \gs_{(cb')}\right)_{;\alpha} \lambda^{(c);\alpha}_\mu \lambda^\mu_{(a)} &= 2 p_{ij} \delta \lambda^i_{(c)} \dot{\lambda} ^{j(c)} \delta_a^0 \delta_{b'}^{0'}, \nonumber \\
	\gs_{(cb')} \left(\lambda^{(c);\alpha}_{\mu;\alpha} \lambda^\mu_{(a)} - \RC_{(a)}^{\ (c)} \right) &= (\ln L^2)\, \ddot{}\ \delta_a^0 \delta_{b'}^{0'}.
\end{align}

Therefore, the solution of the first equation in the chain 
\begin{equation}
	(a^{(1)}_{\mu\tau'} + \sigma^{;\alpha} a^{(1)}_{\mu\tau';\alpha}) \lambda^\mu_{(a)} \lambda^{\tau'}_{(b')}  = \delta_a^0 \delta_{b'}^{0'}  \left(\delta(  \ln L^2)\dot{} + p_{ij} \delta \lambda^i_{(a)} \delta \lambda^{j(a)} \right) \hspace{-.7em} \dot{\phantom{L^2}},
\end{equation}
reads 
\begin{equation}\label{eq:a1}
	a^{(1)}_{\mu\tau'} = \delta_\mu^u \delta_{\tau'}^{u'} \frac{1}{\delta u}  \left(\delta( \ln L^2)\dot{} + p_{ij} \delta \lambda^i_{(a)} \delta \lambda^{j(a)}\right).
\end{equation}
All higher-order terms vanish, $a^{(n)}_{\mu\tau'} =0$ for $n \geq 2$, since $a^{(1)}_{\mu\tau'}$ depends only on $u$ and $u'$, and possesses only a $uu'$ component.

In the coincidence limit$\xb' \to \xb$, we find $[a^{(1)}_{\mu\tau'}] = - \RC_{\mu\tau}$.

This result can also be obtained from the standard formulas for the heat kernel coefficients (see, e.g., \cite{Christensen:1978:Rrcgps,Vassilevich:2003:hkeum}). In taking the coincidence limit, one must account for $p^{ij} \approx \gamma^{ij} \delta u$ and $p_{ij} \approx \gamma_{ij} / \delta u$, together with Eq. \eqref{eq:Rmn}.

The Green function thus takes the form 
\begin{equation}\label{eq:Dover}
	\overline{D}_{\mu\tau'} = \frac{\Delta^{\frac 12}}{8\pi} \gs_{\mu\tau'} \delta (\sigma) +  \frac{\Delta^{\frac 12}}{16\pi } \theta (-\sigma) a^{(1)}_{\mu\tau'},
\end{equation}
and 
\begin{equation}
	h_{\mu\tau'} = \frac{\Delta^{\frac 12}}{8\pi } a^{(1)}_{\mu\tau'},
\end{equation}
where $h_{\mu\tau'}$ is given by Eq. \eqref{eq:tailw}. This result coincides with that obtained in Ref.\,\cite{Kunzle:1968:Mfshp} by an alternative method.

To construct the Feynman Green function, we perform the analytical continuation of the Hadamard elementary solution \cite{Adler:1977:Rstvsppgbm}
\begin{equation}
	D^\Fr_{\mu\tau'} = \frac{\ib \Delta^{\frac{1}{2}}}{(4\pi)^2} \left(\frac{2 \gs_{\mu\tau'}}{\sigma + \ib 0} + v_{\mu\tau'} \ln (\sigma + \ib 0) + w_{\mu\tau'}\right),
\end{equation}
where the bivectors $v_{\mu\tau'},w_{\mu\tau'}$ are expressed as series expansions 
\begin{equation}
	v_{\mu\tau'} = \sum_{n=0}^\infty v^{(n)}_{\mu\tau'} \sigma^n,\ w = \sum_{n=0}^\infty w^{(n)}_{\mu\tau'} \sigma^n.
\end{equation}
The chain of equations for $v^{(n)}_{\mu\tau'}$ yields $v^{(0)}_{\mu\tau'} = - a^{(1)}_{\mu\tau'}$ and $v^{(n)}_{\mu\tau'} = 0$ for $n\geq 1$.

The chain for $w^{(n)}_{\mu\tau'}$ has the form 
\begin{equation}
	(n+1) w^{(n)}_{\mu\tau'} + \sigma^{;\alpha} w^{(n)}_{\mu\tau';\alpha} = - \frac{1}{2n}\left(\Delta^{-\frac 12} (\Delta^{\frac 12} w^{(n-1)}_{\mu\tau'})_{;\alpha}^{\ \ ;\alpha} - \RC_\mu^\alpha w^{(n-1)}_{\alpha\tau'}\right)\label{eq:wn},
\end{equation}
for $n \geq 1$. 

Choosing $w^{(0)}_{\mu\tau'} = 0$ gives $w_{\mu\tau'} = 0$ identically, and the Feynman Green function becomes 
\begin{equation}
	D^\Fr_{\mu\tau'} = \frac{\ib \Delta^{\frac{1}{2}}}{(4\pi)^2} \left(\frac{2 \gs_{\mu\tau'}}{\sigma + \ib 0} - a^{(1)}_{\mu\tau'} \ln (\sigma + \ib 0) \right),
\end{equation}
where the first heat kernel coefficient $a^{(1)}_{\mu\tau'}$ is given by Eq. \eqref{eq:a1}. Taking the real part yields Eq. \eqref{eq:Dover}, while the Hadamard elementary solution is given by 
\begin{equation}
	D^{(1)}_{\mu\tau'} = 2 \im D^\Fr_{\mu\tau'} = \frac{\Delta^{\frac{1}{2}}}{4\pi^2} \left(\frac{\mathcal{P}}{\sigma} \gs_{\mu\tau'} - \frac{1}{2} a^{(1)}_{\mu\tau'} \ln |\sigma | \right).
\end{equation}

As shown by Wald \cite{Wald:1978:Taciqfcs}, in order to compute the energy--momentum tensor one must take $w^{(0)} = 0$ and subtract the term $[a^{(2)}] \gs_{\mu\tau}/64\pi^2$ from the expression for the tensor, where $[\dots]$ denotes the coincidence limit. In the present case, $a^{(2)} =0 $, and therefore no subtraction is required.

\section{Discussion and conclusion} \label{sec:conclusion}

In the preceding sections, we have rederived the scalar Green function first obtained in Ref.\,\cite{Gibbons:1975:Qfpps} and computed the massless vector Green function. Three distinct computational methods were employed: (i) direct solution of the differential equation for the Green function using the complete set of eigenfunctions, (ii) construction via the DeWitt recursive scheme, and (iii) construction via the Hadamard recursive scheme. In all cases considered, the resulting Green functions coincide with their DeWitt--Schwinger form.

Within the point-splitting approach, the vacuum expectation value of the renormalized energy--momentum tensor (EMT) for the electromagnetic and massless scalar fields may be obtained through the procedure 
\begin{align}
	\langle T_\mu^{\ \nu} \rangle^\ren & = \lim_{x' \to x} \left[ \nabla^{\alpha'}_\alpha G^{(1)}_\ren{}_\mu^{\ \nu'} - \nabla_\mu^{\alpha'} G^{(1)}_\ren{}_\alpha^{\ \nu'} - \nabla_\alpha^{\nu'} G^{(1)}_\ren{}_\mu^{\ \alpha'} + \nabla_\mu^{\nu'} G^{(1)}_\ren{}_\alpha^{\ \alpha'}\right.\nonumber \\
	&  \left. - \tfrac{1}{2} \gs_\mu^{\ \nu'} \left(\nabla_\alpha^{\ \alpha'} G^{(1)}_\ren{}_\beta^{\ \beta'} - \nabla_\alpha^{\ \beta'} G^{(1)}_\ren{}_\beta^{\ \alpha'}\right)\right],\nonumber \\
	\langle T_\mu^{\ \nu} \rangle^\ren & = \lim_{x' \to x} \left[ (1-2\xi) \nabla_{\mu\nu'} + \gs_{\mu\nu'} \left(2\xi - \tfrac{1}{2}\right) \gs^{\alpha\beta'} \nabla_{\alpha\beta'}\right.\nonumber\\
	&\left. - \xi \left(\gs_{\mu\rho'} \nabla^{\rho'}_{\ \nu'} + \gs_{\nu'\sigma} \nabla^\sigma_{\ \mu}\right)\right]G^{(1)}_\ren,
\end{align}
respectively. Here, $\xi$ denotes the non--minimal coupling, while $G^{(1)}_\ren{}_\nu^{\ \nu'}$ and $G^{(1)}_\ren$ represent the renormalized Hadamard elementary solutions. The renormalization consists of subtracting the first three terms of the DeWitt--Schwinger expansion (up to the second heat-kernel coefficient). Consequently, $G^{(1)}_\ren{}_\nu^{\ \nu'} =0$ and $G^{(1)}_\ren =0$, and the vacuum expectation value of the EMT vanishes exactly, implying the absence of particle production for both scalar and vector fields in a plane gravitational wave background. For scalar particles, this result was first obtained by Gibbons \cite{Gibbons:1975:Qfpps} and later confirmed by Garriga and Verdaguer \cite{Garriga:1991:Sqpgpwa} using the Pauli--Villars renormalization prescription. The same conclusion may be easily obtained from Ref.\,\cite{Boasso:2025:Neapcd} within the framework of the effective action approach.

Consider now a flux of photons propagating along the $x$-direction. Since $a_2 = 0$, there is no conformal anomaly \cite{Birrell:1982:Qfcs}, and the trace of the renormalized EMT vanishes. The EMT therefore shares the same properties as that of an electromagnetic plane wave propagating in the $x$-direction. It is well known \cite{Landau:1975:CTF} that this tensor possesses only three nonzero components: $T^{tt}_\ren = T^{tx}_\ren = T^{xx}_\ren = W$, where $W = W(u)$ is the energy density. In null coordinates $(u, v)$, the EMT has a single nonzero component, $T_{uu} = 2W(u)$.

At the same time, Refs.\,\cite{Jones:2017:Ppgwb,Jones:2018:Sfvevigwb,Jones:2019:Igrer} report that massless particle production occurs within the framework of the Bogolyubov coefficient method. In this picture, the flux of created particles is strongly aligned -- both in direction and velocity -- with the propagation of the gravitational wave.

This conclusion is based on the observation by Garriga and Verdaguer \cite{Garriga:1991:Sqpgpwa} that the Bogolyubov coefficients satisfy $\beta \sim \delta (k_v + l_v)$ and $\alpha \sim \delta (k_v - l_v)$, where $k_v$ and $l_v$ are the $v$-components of the \textit{in} and \textit{out} wave vectors of the scalar field, respectively. For a gravitational wave "sandwich"\/ $u_1 < u_2 < 0$, the matrix $\mathbf{P}$ after passage of the GW packet is given by $\mathbf{P} = \mathbf{A} + u \mathbf{B}$, and, as shown in Ref.\,\cite{Garriga:1991:Sqpgpwa}, 
\begin{align}
	\alpha_{k,l} & = -\frac{\delta(k_v - l_v)}{4\pi \ii k_v\sqrt{|\det \mathbf{B}|}} e^{\frac{\ii}{4k_v} \mathbf{J}^T \left(\mathbf{A}^T \mathbf{B}\right)^{-1}\mathbf{J}}, \nonumber \\
	\beta_{k,l} & =  -\frac{\delta(k_v + l_v)}{4\pi \ii k_v\sqrt{|\det \mathbf{B}|}} e^{-\frac{\ii}{4k_v} \mathbf{I}^T \left(\mathbf{A}^T \mathbf{B}\right)^{-1}\mathbf{I}},
\end{align} 
where $J_j = l_i A^i_j - k_j$ and  $I_j = l_i A^i_j + k_j$. These relations follow directly from the existence of the Killing isotropic vector $\bm{\xi}_1 = \partial_v$. For massive particles, $k_v + l_v >0$, and thus $\beta = 0$, implying no particle production. For massless particles propagating in the direction of the GW, $k_i=l_i=0$, so that $k_v = l_v = 0$. In this case, the arguments of the delta functions in $\beta$ and $\alpha$ vanish, leading to the conclusion that particle production occurs.

This results in an apparent contradiction between the two approaches to particle production calculations. The standard method based on the Green-function formalism \cite{Birrell:1982:Qfcs} predicts no particle creation, whereas the Bogolyubov approach suggests a flux of photons in the direction of the GW. 

The aim of this paper is to draw the attention of the physics community to this unresolved issue. Now, let's make some observations. One possible solution to the problem is the ambiguity of the vacuum state in a curved background, which was first noted by Fulling in Ref.\,\cite{Fulling:1973:NCFQRS}. For the Schwarzschild black hole background, three different vacuums were considered, as noted in the introduction. The notion of a "particle"\/ depends on the choice of "time." Due to the invariance of general relativity with respect to coordinate transformations, any time-like variable can be used as "time"\/ to define particles in relation to it. In the case of plane GW, the natural time coordinate is the retarded variable $u=t-x$, which Gibbons used to calculate the scalar Green function in Ref.\,\cite{Gibbons:1975:Qfpps} and which is used here for the vector Green function. The choice of this variable as "time"\/ is related to the fact that the domain of the GW "sandwich"\/ is described by the interval of this variable, $u\in (u_1,u_2)$. Outside this interval, the spacetime is flat, and the vacuum state is well-defined. However, nothing prevents us from making a different choice of "time."   

It is also ambiguity for Green function itself. In general, a Green function is defined only up to a solution of the associated homogeneous equation, with the homogeneous part determined by the boundary conditions. A well-known example is the electrostatic potential of a point charge at rest in a Schwarzschild spacetime: Copson \cite{Copson:1928:egf} obtained the potential (the three-dimensional Green function) almost a century ago; however, as noted by Linet \cite{Linet:1976:Emsm}, its asymptotic form does not match the Coulomb potential far from the black hole, as it should. The correct behaviour is recovered by adding an appropriate homogeneous solution. In the present case, the fields are free, so no additional homogeneous solution is required, and consequently there is no mechanism to describe particle production.

Another observation is that the velocity of GW coincides with the velocity of light. Therefore, if photons were created inside the sandwich only in the direction of the GW, as affirmed in Refs.\,\cite{Jones:2017:Ppgwb,Jones:2018:Sfvevigwb,Jones:2019:Igrer}, they would remain there and leave with the sandwich. After passing the GW package, we cannot detect the created photons. The created particles can reveal themselves  through a backreaction to the metric.     

\section*{Acknowledgments}
NK is grateful to S. Fulling, D. Fursaev, G. Gibbons and A. Zelnikov for fruitful discussions. The work was supported in part by the grants 2021/10128-0,  2024/22940-0, 2025/13673-0 of S\~ao Paulo Research Foundation (FAPESP).

\appendix

\section{Calculation of the bivector $\gs_{\mu\nu'}$} \label{sec:appA} 

Let us consider the equation $u^\tau \omega_{\mu; \tau}=0$ of parallel transport for a vector $\omega_\mu$ alongside the geodesic line between events $\xb = (t,x,y,z)$ and $\xb' = (t',x',y',z')$. Then, the frame components $\omega_{(a)} = \omega_\mu \lambda^\mu_{(a)}$ obey to the following equations 
\begin{equation}
\frac{\dd \omega_{(a)}}{\dd s} = \Lambda_a^b \omega_{(b)} = u^\tau  \lambda^\mu_{(a);\tau} \lambda_\mu^{(b)}\omega_{(b)}.
\end{equation}

Straightforward calculations give 
\begin{equation}
\Lambda_2^3 = - \Lambda_3^2 =u_v\left( \frac{\gamma_{yz} \dot{\gamma}_{yy} - \dot{\gamma}_{yz} \gamma_{yy}}{2\gamma_{yy} L^2} - \dot{\psi}\right). 
\end{equation} 
Taking $\psi$ in the form   
\begin{equation}
\psi = \int^u_{u'} \frac{\gamma_{yz} \dot{\gamma}_{yy} - \dot{\gamma}_{yz} \gamma_{yy}}{2\gamma_{yy} L^2}du,
\end{equation}
we arrive with equations 
\begin{gather}
-u_v \dot{\omega}_{(0)} = \Lambda_2^1 \omega_{(2)} + \Lambda_3^1 \omega_{(3)},\ \dot{\omega}_{(1)} = 0, \nonumber\\
-u_v \dot{\omega}_{(2)} = \Lambda_2^1 \omega_{(1)}, \ -u_v \dot{\omega}_{(3)} = \Lambda_3^1 \omega_{(1)}.  
\end{gather}

The coefficients have the following form 
\begin{equation}
\Lambda_2^1 = - u_v \left(\upsilon_\mu \lambda^\mu_{(2)}\right)\hspace{-.7em}\dot{\phantom{L^2}},\ \Lambda_3^1 = - u_v \left(\upsilon_\mu \lambda^\mu_{(3)}\right)\hspace{-.7em}\dot{\phantom{L^2}},
\end{equation}
where $\upsilon_\mu = \left(0,0,u_y^0/u_v^0,u_z^0/u_v^0\right)$ is the constant along geodesic line vector. Solution of this system reads
\begin{gather}
\delta \omega_{(1)} = 0,\ \delta\omega_{(2)} = \omega_{(1)} \delta \upsilon_{(2)},\ \delta\omega_{(3)} = \omega_{(1)} \delta \upsilon_{(3)}, \nonumber \\
\delta\omega_{(0)} = \delta \upsilon_{(a)} \omega^{(a)}(u')   + \frac{1}{2} \delta \upsilon_{(a)} \delta \upsilon^{(a)} \omega_{(1)} ,
\end{gather}
where $\delta \upsilon_{(a)} = \upsilon_\mu \delta \lambda^\mu_{(a)}$ and $\delta \omega_{(a)} = \omega_{(a)}(u) - \omega_{(a)}(u')$. 

Now, we represent the solution in the form $\omega_{(a)}(u) = \gs_{(a)(b)} \omega^{(b)}(u')$:
\begin{align}
\omega_{(0)}(u) &= - \omega^{(1)}(u')  + \delta \upsilon_{(a)} \omega^{(a)}(u')   - \frac{1}{2} \delta \upsilon_{(a)} \delta \upsilon^{(a)} \omega^{(0)}, \nonumber \\
\omega_{(2)}(u) &= \omega^{(2)}(u')  -  \delta \upsilon_{(2)}\omega^{(0)}, \nonumber \\ 
\omega_{(3)}(u) &= \omega^{(3)}(u')  -  \delta \upsilon_{(3)}\omega^{(0)}, \nonumber \\ 
\omega_{(1)}(u) &= - \omega^{(0)}(u'). 
\end{align}
From these relations we obtain bivector of parallel transport along geodesic line \eqref{eq:gab}. Integrating the first integrals of geodesic $u_y = u_y^0$, $u_z = u_z^0$ we obtain that $u_i^0/u_v^0 = - p_{ij} \delta x^j$. 

%

\end{document}